# How *enlightened self-interest* guided global vaccine sharing benefits all: a modelling study

Zhenyu Han[1,2] (https://orcid.org/0000-0001-9634-7962), Lin Chen[2] (https://orcid.org/0000-0002-2605-749X), Qianyue Hao[1,2](https://orcid.org/0000-0002-7109-3588), Qiwei He[3,4] (https://orcid.org/0000-0003-1410-3839), Katherine Budeski[5], Depeng Jin[1,2], Fengli Xu[2]†(https://orcid.org/0000-0002-5720-4026), Kun Tang[3,4]† (https://orcid.org/0000-0002-5444-186X), Yong Li[1,2]† (https://orcid.org/0000-0001-5617-1659)

[1] Beijing National Research Center for Information Science and Technology (BNRist), Beijing, P. R. China

[2] Department of Electronic Engineering, Tsinghua University, Beijing, P. R. China

[3] Vanke School of Public Health, Tsinghua University, Beijing, P. R. China

[4] Institute for Healthy China, Tsinghua University, Beijing, P. R. China

[5] Nuffield Department of Medicine, University of Oxford, UK

†Joint senior authorship.

ABSTRACT

**Background** Despite the consensus that vaccines play an important role in combating the global spread of infectious diseases, vaccine inequity is still rampant with deep-seated mentality of self-priority. This study aims to evaluate the existence and possible outcomes of a more equitable global vaccine distribution and explore a concrete incentive mechanism that promotes vaccine equity.

**Methods** We design a metapopulation epidemiological model that simultaneously considers global vaccine distribution and human mobility, which is then calibrated by the number of infections and real-world vaccination records during COVID-19 pandemic from March 2020 to July 2021. We explore the possibility of the *enlightened self-interest* incentive mechanism, *i.e.*, improving one's own epidemic outcomes by sharing vaccines with other countries, by evaluating the number of infections and deaths under various vaccine sharing strategies using the proposed model. To understand how these strategies affect the national interests, we distinguish the imported and local cases for further cost-benefit analyses that rationalize the *enlightened self-interest* incentive mechanism behind vaccine sharing.

**Results** The proposed model accurately reproduces the real-world cumulative infections for both global and regional epidemics with $R^2 > 0.990$, which can support the following evaluations of different vaccine sharing strategies. High-income countries can reduce 16.7 (95% CI: 8.4-24.9, $P<0.001$) million infection cases and 82.0 (95% CI: 76.6-87.4, $P<0.001$) thousand deaths on average by more actively sharing vaccines in an *enlightened self-interest* manner, where the reduced internationally imported cases outweigh the threat from increased local infections. Such vaccine sharing strategies can also reduce 4.3 (95% CI: 1.2-7.5, $P<0.01$) million infections and 7.0 (95% CI: 5.7-8.3, $P<0.001$) thousand deaths in middle- and low-income countries, achieving a win-win situation for the globe. Lastly, the more equitable vaccine distribution can largely avoid the global mobility reduction needed for pandemic control.

**Conclusions** We study the incentive mechanism of *enlightened self-interest* that can motivate vaccine equity by realigning the national interest to more equitable vaccine distributions. The positive results could promote multilateral collaborations in global vaccine redistribution and reconcile conflicted national interests, which facilitates a mutually benefiting world.

## INTRODUCTION

Vaccines continue to be the most crucial resource for curbing the global spread of infectious diseases [1], achieving effective epidemic control at a relatively low economic cost compared with lockdown or quarantine measures. The scarcity of vaccine resources during the early stage of disease outbreak makes them a sought-after commodity, hoarded by high-income countries under the influence of self-priority policies. For instance, during the combat against the COVID-19 pandemic, more than three-fourths of the 7.8 billion manufactured vaccine doses are possessed by a few high-income countries as of 27 November 2021 [2], which is "not a supply problem, but an allocation problem" urged by World Health Organization (WHO). The inequity of vaccine resources not only creates a hotbed for disease transmission in the low- and middle-income countries, but also undermines epidemic control in high-income countries in this highly intertwined and globalized world [3-5].

To remedy the inequitable vaccine distribution, international organizations are taking action. A global risk-sharing mechanism for pooled procurement and equitable redistribution of vaccines was proposed by the COVID-19 Vaccines Global Access (COVAX) Facility in April 2020, which purchases vaccine doses in bulk at a discount and distributes equitably among the participants [6,7]. Although it aims ambitiously to provide at least 2 billion vaccine doses to cover 20% of each country's population in 2021 [7], it has been repeatedly falling behind its goals and only delivered less than half of the target doses by the beginning of 2022 [8,9]. One crucial reason for its failure is the absence of clear incentives for high-income participants to promote global vaccine sharing with their own wealth. According to the rational choice theory, the decisions made by individuals will maximize their own utility considering the corresponding costs and benefits [10]. Although it is ethically recognized that more equitable vaccine distribution is in "everyone's best interest" [11,12], a practical incentive with quantitative, real-world data supported evaluation is an important yet missing piece for promoting global vaccine equity [13-15].

Motivated by the importance vaccine equity, many scientists and public health researchers are dedicated to optimizing the distribution of vaccine doses and understanding its consequences. A large body of recent works focused on improving the effectiveness [16-18] and equity [19-23] of the domestic distribution of vaccines. These studies are incapable of modelling the national interests of sovereign countries and scaling up to global vaccine distribution [11,24]. To explore the effect of global vaccine sharing, several early attempts employed highly simplified models with hypothetical parameters, which usually overlook the structure of global mobility [20] and are lack of real-world data supported evidence [14,15,25,26]. Although these works provided important theoretical results for various hypothetical scenarios, they cannot adequately reflect the complex situation of real-world pandemics, which limited their capability of promoting global vaccine sharing with practical incentives.

In this study, we aim to investigate the incentive mechanism of *enlightened self-interest* for global vaccine sharing through quantitative modelling analyses, where each party's willingness to share a portion of vaccines to assist others in return promotes their own interest. Specifically, we propose a novel epidemiological model that provides a connected perspective of the world by simultaneously considering global vaccine distribution and human mobility. Leveraging the proposed model, we evaluate a wide range of vaccine sharing strategies to explore the feasibility of *enlightened self-interest* phenomenon, which has the potential to substantially motivate global cooperation in combating infectious diseases with equitable vaccine access. We assume the shared vaccines are pooled globally and redistributed to all regions in proportion to their population sizes as COVAX, based on which *enlightened self-interest* can be achieved through balancing the two competing influence pathways of vaccine sharing and human mobility. The objective of this study is to propose and rationalize the practical incentive mechanism of *enlightened self-interest* that facilitates more equitable global vaccine distribution through a quantitative, real-world data supported modelling approach.

## METHODS

We conduct a modelling study that extends the classic epidemiological model (*e.g.*, SIR model [27] ) by considering regional mobility flows and vaccination process. Leveraging the calibrated model on real-world infection records, we explore several vaccine sharing scenarios and the corresponding outcomes in terms of averted infections and deaths, aligned national interests with more equitable vaccine distribution. We use one-way Analysis of Variance (ANOVA) to validate the statistical significance among the contrasting scenarios.

### Regional division and scenario setup

In this study, we take the COVID-19 pandemic as an example to evaluate the existence and possible outcomes of a more equitable global vaccine distribution with rich real-world data support. Considering the varying epidemic situation and abilities to access vaccine supplies across the world, we divide the globe into seven regions according to the World Bank definitions [28] in **Figure 1,** panel A: East Asia & Pacific (EAS), Europe & Central Asia (ECS), Latin America & Caribbean (LCN), Middle East & North Africa (MEA), North America (NAC), South Asia (SAS), and Sub-Saharan Africa (SSF) (see Supplementary **Table S1, S2** in the Online Supplementary Document). We divide these regions into two groups based on their production capability (see Supplementary Method in the Online Supplementary Document) [29,30]: EAS, ECS, NAC and SAS are noted as *vaccine-producing* *regions*. Other regions that lack manufacturing capability (LCN, MEA and SSF) are referred to as *non-vaccine-producing regions*. Since high-income countries have enough resources to secure their vaccine production capability, we observe the vaccine-producing regions mainly contain high-income countries, while lower-middle-income countries mostly are located in non-vaccine-producing regions (see Supplementary Table S3 in the **Online Supplementary Document**).

### Global mobility-aware epidemiological model design

To capture the epidemic dynamics under different vaccine sharing strategies, we propose a novel epidemiological model that systematically accounts for the breakthrough infection during vaccination process and the global mobility network. Specifically, each region maintains its own compartmentalized states for susceptible (*S*), vaccinated (*V*), infectious (*I*), recovered (*R*), and deceased (*D*) people. To faithfully reproduce real-world COVID-19 spreading with these compartmentalized states and avoid the strong assumption of once-and-for-all vaccination [15], we consider infections as a probabilistic process where the vaccinated population has a reduced infection probability according to the up-to-date medical research [31-33], *i.e.*, the breakthrough infection. We argue that to faithfully model the epidemic outcomes under different vaccine sharing strategies without overestimating their effectiveness, breakthrough infection is necessary for the model, where both susceptible people and part of vaccinated people (considering the non-100% vaccine effectiveness) can get infected, which generates $I_s$ and $I_v$ accordingly. In our model, the vaccinated infection cases tend to have a lower death rate [31-33], which affects the transfer possibility from infectious to recovered and deceased. The above contagion process in each region is characterized by the following ordinary differential equations:

$$\frac{dS_n}{dt} = -vcc(n,t) - \beta_n S_n (I_{s,n} + I_{v,n}),$$

$$\frac{dV_n}{dt} = vcc(n,t) - \beta_n (1-\kappa) V_n (I_{s,n} + I_{v,n}),$$

$$\frac{dI_{s,n}}{dt} = \beta_n S_n (I_{s,n} + I_{v,n}) - \gamma_n I_{s,n} - \psi I_{s,n},$$

$$\frac{dI_{v,n}}{dt} = \beta_n(1-\kappa)V_n(I_{s,n}+I_{v,n}) - \gamma_n I_{v,n} - \psi(1-\sigma)I_{v,n},$$

$$\frac{dR_{s,n}}{dt} = \gamma_n I_{s,n},$$

$$\frac{dR_{v,n}}{dt} = \gamma_n I_{v,n},$$

$$\frac{dD_{s,n}}{dt} = \psi I_{s,n},$$

$$\frac{dD_{v,n}}{dt} = \psi(1-\sigma)I_{v,n},$$

where $S_n, V_n, I_{\star,n}, R_{\star,n}, D_{\star,n}$ are the susceptible, vaccinated, infected, recovered and deceased people in region $n$. $vcc(n,t)$ is the number of people who can be vaccinated in region $n$ at time $t$ derived by various vaccine sharing strategies. Specifically, we divide the infected, recovered and deceased people according to their vaccination status into $I_{s,n}, R_{s,n}, D_{s,n}$ and $I_{v,n}, R_{v,n}, D_{v,n}$ for unvaccinated and vaccinated people accordingly. It enables us to explicitly model the breakthrough infection process.

Based on these regional models, we overlay a mobility network connecting these regions to form a global view of the epidemic. We adopt the international air traffic records from OAG, which has been proven to be highly informative for epidemic modelling [34]. Leveraging this air traffic records, we reconstruct the global mobility network with all transportations. Considering the privacy issue, the detailed health status of passengers cannot be acquired. The continued existence of asymptomatic COVID-19 cases makes it harder to distinguish the flow of infection cases [35]. Under this circumstance, we assume infection cases move across regions proportionally to the prevalence of the source region. This assumption maximizes the utility of real-world mobility data, while also providing us the ability to simulate the epidemic dynamics under different vaccine sharing strategies. Detailed mobility network construction can be found in Supplementary Method in the **Online Supplementary Document**.

### Characterizing different vaccine sharing strategies

We define two parameters to characterize different vaccine sharing strategies: the *start-sharing point* and the *sharing percentage*. Each vaccine-producing region starts with only vaccinating its citizens until its vaccination rate reaches the *start-sharing point*. Then it starts to share the newly acquired vaccines by the *sharing percentage*. The shared vaccines are pooled globally and redistributed to all regions in proportion to their population sizes, which is consistent with the COVAX framework [36]. As such, we can design a series of typical vaccine sharing strategies by formulating different combinations of the *start-sharing point* (*a*) and the *sharing percentage* (*b*). For example, when *b*=0%, vaccine-producing regions will not share any vaccine even after all their citizens are fully vaccinated, which we refer to as the *non-sharing* strategy. When *a*=100%, *b*=100%, the vaccine-producing regions will share all their newly produced vaccines after finishing their own citizens, which is denoted as the *selfish* strategy. Both the *non-sharing* and *selfish* strategies indicate the vaccine-producing region will prioritize their own citizens for full vaccination before they start sharing, which represents the typical self-priority behaviours. On the contrary, an *altruistic* strategy will share 100% of their vaccine supplies from the very beginning, where *a*=0%, *b*=100%. More fine-grained vaccine sharing strategies can be defined with specific combinations of *a* and *b*, ranging from the *non-sharing* to the *altruistic* strategy.

We evaluate the vaccine sharing strategies with different parameters and compare the corresponding number of infections or deaths in each region with the *selfish* strategy. The feasible parameter combinations for *enlightened self-interest* guided vaccine sharing are defined as those simultaneously reducing the infections in both vaccine-producing regions and non-vaccine-producing regions.

### Disentangling imported transmission and local transmission

New cases can be divided into two types: local transmission caused by the initial cases in the region, and imported transmission caused by the inflow of infection cases and their descendant cases. To investigate how vaccine sharing strategies affect the epidemic outcomes in vaccine-producing regions and non-vaccine-producing regions, we need to disentangle these two types of transmission. In this work, we define the imported transmission by summing up the number of imported infections and their subsequent infections, and define the local transmission as infections caused by pre-existing local cases. We trace back the above process until the COVID-19 vaccines are available, which is in December 2020. Detailed calculation process and illustration can be found in Supplementary Method and Supplementary **Figure S5** in the Online Supplementary Document.

### Quantifying the impact of vaccine sharing on inter-regional mobility policy

We measure the mobility benefit of different vaccine sharing strategies as the equivalent mobility reduction under the *selfish* strategy. That is, we calculate the required reduction of inter-regional mobility under the *selfish* strategy in order to achieve a similar level of infection prevention as the inspected vaccine sharing strategies with no mobility constraint, which is considered as the equivalent mobility benefit of the given vaccine sharing strategy.

## RESULTS

### Reproducing epidemic developments under real-world vaccine distribution

We first showcase the significant global mismatch between COVID-19 prevalence, measured by seroprevalence statistics, and vaccine distribution in different regions (see **Figure 1,** panel A, B). Regions with high seroprevalence possess few vaccine supplies, where nearly ten people share a single vaccine in SSF. By contrast, vaccine-producing regions such as EAS and NAC have the largest vaccine supplies, albeit with relatively lower seroprevalence. The mismatch between seroprevalence and vaccine distribution demonstrates the substantial vaccine inequity around the world, suggesting the great potential for more equitable vaccine sharing.

To capture the above global mismatch and evaluate the epidemic outcomes under more equitable vaccine sharing strategies, we propose a quantitative, real-world data supported modelling approach. Leveraging a metapopulation schema with compartmentalized states for susceptible ($S$), vaccinated ($V$), infectious ($I$), recovered ($R$), and deceased ($D$) people, we model the world into several regions that are interconnected by both vaccination sharing and population flows. It captures both the spatial and temporal heterogeneity of the epidemiological severity and vaccine resources (see Methods and Supplementary Methods for details in the **Online Supplementary Document**). The proposed model accurately reproduces the vastly different and ever-changing epidemic dynamics in all the regions with $R^2 > 0.990$ (**Figure 1,** panel C, blue lines), which demonstrates sufficient capability to capture the disease transmission processes in a connected world.

### *Enlightened self-interest* guided vaccine sharing strategies

As an intuitive example, we try to reduce the mismatch between COVID-19 prevalence and vaccine distributions through *altruistic* strategy, where vaccine-producing regions will share all of their vaccine supplies to the global vaccine pool from the beginning. This naïve policy will lead to a reduction of 10.77% (95% CI: 10.38%-11.17%, $P<0.001$) global infections compared to the *selfish* strategy (see Supplementary **Fig S1** in the **Online Supplementary Document**), which is a significant improvement in terms of global epidemic outcome, yet at the cost of incurring more infections in vaccine-producing regions, *e.g.*, ECS suffers 4.37% (95% CI: 3.97%-4.74%, $P<0.001$) more infections.

We argue that such trade-offs can be avoided with more deliberated vaccine sharing strategies. As shown in **Figure 2,** panel A, we demonstrate the likelihood to achieve such *enlightened self-interest* of each parameter combination by different colours. We find when the *start-sharing point* reaches 60%, there are possibilities to achieve *enlightened self-interest*, as the dotted line in **Figure 2,** panel A shows. Besides, at a higher *start-sharing point*, vaccine-sharing strategies can share more vaccine supplies with other regions. As a conservative estimation, we also provide the parameter combinations that achieve *enlightened self-interest* over 80% likelihood, as the dashed line in **Figure 2,** panel A shows. Under this strict definition, we still observe that when the local vaccination rate reaches around 80%, vaccine-producing regions can still share most of their vaccine supplies with other regions. Under *enlightened self-interest* guided vaccine sharing strategies, sharing vaccines not only benefits the non-vaccine-producing regions but also benefits the vaccine-producing regions themselves. Detailed epidemic outcomes for each region are available in Supplementary **Fig S2** in the **Online Supplementary Document**.

We further evaluate how different levels of sharing, represented by the parameter *sharing percentage* (*b*), affect the infections and deaths. Specifically, we search the optimal *start-sharing point* for different *sharing percentages* to achieve their best epidemic outcomes in the feasible parameter combinations of *enlightened self-interest* guided strategies, noted as *b*%-Sharing accordingly. In **Figure 2,** panel B, we observe a higher *sharing percentage* will result in a greater reduction in both infections and deaths for all kinds of regions. As an example, the *100%-sharing* strategy reduces 16.7 (95% CI: 8.4-24.9, *P*<0.001) million infections in vaccine-producing regions and 4.3 (95% CI: 1.2-7.5, *P*<0.01) million in non-vaccine-producing regions. It also reduces 82.0 (95% CI: 76.6-87.4, *P*<0.001) thousand infections and 7.0 (95% CI: 5.7-8.3, *P*<0.001) thousand deaths in vaccine-producing regions and non-vaccine-producing regions, respectively. Moreover, 79.5% (95% CI: 76.9%-87.5%) of globally reduced infections belong to vaccine-producing regions, which provides a strong incentive for proactively sharing vaccines. On the other side, the *non-sharing* strategy guided by extreme self-priority results in considerably more infections in every region. In EAS that receives the greatest impact, we observe 14.8 (95% CI: 13.3-16.4, *P*<0.001) million extra infections.

Furthermore, we compare the geographical distributions of vaccines under the *100%-sharing* strategy and the *selfish* strategy. Under the *selfish* strategy (**Figure 2,** panel C left), non-vaccine-producing regions, such as LCN and SSF, possess few vaccine supplies. By contrast, under *100%-sharing* strategy (**Figure 2,** panel C right), non-vaccine producing regions receive vaccines without bringing great vaccine loss in vaccine-producing regions, which reduces the significant inequity in vaccine distribution. From the above, we demonstrate the feasibility of vaccine sharing strategies guided by the *enlightened self-interest* incentive mechanism, which can effectively relieve the epidemic outcomes for both vaccine-producing and non-vaccine-producing regions.

## Rationalizing the *enlightened self-interest* incentive mechanism

Having shown the feasibility of vaccine sharing strategies that benefit all regions, we further explore how such mutual benefits are achieved. The epidemic dynamics of different regions are coupled via two competing influence pathways: vaccine sharing and inter-regional mobility (**Figure 3,** panel A). As for vaccine sharing, vaccine-producing regions will share their vaccines with non-vaccine-producing regions, which can improve the global vaccine equity but might also increase their own local epidemic risk. However, due to the second pathway of highly intertwined inter-regional mobility, vaccine sharing might in return also benefit the vaccine-producing regions.

To validate the above hypothesis, first we evaluate the marginal utility of vaccines, *i.e.*, the averted cases per extra vaccine. As shown in **Figure 3,** panel B, the marginal utility of vaccines is significantly higher in non-vaccine-producing regions due to the severe shortage, where, for example, the averted cases per vaccine in SSF is 79.2 (95% CI: 58.2-100.1, *P*<0.001) times of NAC. In fact, we observe a strong negative correlation (Spearman R = -0.429) between the vaccination rate and the averted cases per extra vaccine (see Supplementary Method in the Online Supplementary Document). Therefore, due to the inter-regional mobility, the higher marginal utility of vaccines in non-vaccine-producing regions indicates that vaccine sharing could reduce imported cases in return for vaccine-producing regions by maximizing the potential value of limited vaccine supplies, achieving *enlightened self-interest*.

We evaluate how the second influence pathway, *i.e.*, the inter-regional mobility, affects vaccine-producing regions. Specifically, we simulate the flow of reduced infection cases under the *enlightened self-interest* guided sharing strategy compared with the *selfish* strategy in **Figure 3,** panel C**.** We observe a significant reduction of imported cases in ECS and NAC, which accounts for 49.7% (95%CI: 48.8%-50.6%, *P*<0.001) in all regions. Such benefits for vaccine-producing regions are mainly achieved from reduced infection cases in EAS, SAS, SSF and MEA. Except for EAS, where its large population base contributes to the reduction of infected inter-regional travellers, these regions have relatively fewer vaccines per capita, which validates the findings in **Figure 3,** panel B. The significantly reduced imported cases in vaccine-producing regions compensate for or even benefit their vaccine sharing behaviours (see Supplementary **Figure S3, S4** in the Online Supplementary Document). These results suggest the disproportionate COVID-19 vaccine effect and strong inter-regional mobility coupling combined to facilitate the vaccine sharing strategies that meet the vaccine-producing regions' *enlightened self-interests*, where the vaccine-producing regions are in fact better off by sharing vaccines globally instead of adopting the *selfish* strategy.

To further quantify the trade-off between the competing influence pathways of vaccine sharing and inter-regional mobility, we disentangle the averted local transmission and imported transmission with respect to the *selfish* strategy as the measurements for our proposed *enlightened self-interest* guided vaccine sharing strategy (see Method for details). As shown in **Figure 3,** panel D, all the regions have a net positive benefit in imported transmission since the global COVID-19 infection is reduced by vaccine sharing. Moreover, the vaccine-producing regions, *i.e.*, ECS, NAC, EAS, and SAS, receive the most benefit from reduced imported transmission as the main destination of global mobility flow (see Supplementary **Figure S3** in the Online Supplementary Document). Besides, for the non-vaccine-producing regions in Group I, *i.e.*, LCN, MEA and SSA, they receive more benefit from local transmission than imported transmission under the *enlightened self-interest* guided vaccine sharing strategy, because they gain access to more globally pooled vaccines. Group II are the regions where both the imported and local transmission are reduced substantially, which consists of the regions with large population and relatively small vaccine production capabilities, *i.e.*, EAS and SAS. Therefore, they actually receive more vaccines from the global pool albeit being vaccine-producing regions. As a result, the averted case in local transmission is 6.65 (95% CI: 5.28-9.02, *P*<0.001) and 15.1 (95% CI: 13.7-16.5, *P*<0.001) times of the imported transmission accordingly. For group III of NAC and ECS, the change in local transmission is relatively small, because they are the main regions that give out extra vaccines with low margin utility. However, the benefit of reduced imported transmission outweighs the slight increase in local transmission by over 32.9 (95% CI: 27.5-38.2, *P*<0.001) times, which aligns vaccine sharing with their *enlightened self-interest*. Therefore, we reveal a feasible incentive mechanism to promote equitable global vaccine redistribution driven by *enlightened self-interest*.

## Equitable vaccine sharing promotes a connected world

During the pandemic, most parts of the world have implemented various travel restriction policies, which significantly reduced the inter-regional mobility by approximately 80.7% (measured by the number of travellers). As another incentive for more equitable vaccine distribution, we examine how active vaccine sharing promotes global connectivity without worsening the epidemic outcomes.

Intuitively, a more equitable global vaccine distribution can reduce the infection risk embedded in the inter-regional population flow. **Figure 4,** panel A shows that a more *altruistic* sharing strategy, *i.e.*, when the vaccine sharing starts earlier and the sharing percentage is larger, can facilitate a higher level of inter-regional mobility, which can prevent up to 94.8% (95% CI: 94.3%-95.3%, *P*<0.001) of mobility reduction.

Moreover, a major portion of the averted mobility reduction is associated with the vaccine-producing regions, as shown in **Figure 4,** panel B. NAC and ECS account for 48.8% (95% CI: 48.3%-49.3%, *P*<0.001) of the total prevented mobility inflow reduction around the world, and EAS also accounts for 15.3% (95% CI: 15.2%-15.5%, *P*<0.001) of prevented mobility inflow reduction due to its large population. We further validate this finding across different sharing strategies in **Figure 4,** panel C, where vaccine-producing regions constantly benefit more than non-vaccine-producing regions. Major vaccine-producing regions, *i.e.*, NAC and ECS, benefit most under different *enlightened self-interest* guided vaccine sharing strategies in terms of averted mobility reduction. Specifically, more than 3.21 (95% CI: 3.16-3.27, *P*<0.001) billion inter-regional travellers can move freely under *100%-sharing* strategies to NAC and ECS. These observations on global mobility provide yet another incentive for vaccine-producing regions to actively share vaccine supplies with the world, which suggests they can also benefit more in the transition back to a more connected world.

# DISCUSSION

In this study, we propose a novel epidemiological model that can simultaneously model global vaccine distribution and real-world human mobility, which serves as a platform for revealing how different vaccine sharing strategies affect the connected world. We quantitatively demonstrate the existence and potential value of *enlightened self-interest* incentive mechanism during COVID-19 pandemic, which can promote global vaccine sharing without harming the interest of vaccine-producing regions.

## Principal Findings

Our epidemiological model goes beyond existing research that adopts a once-and-for-all vaccination assumption by explicitly considering the breakthrough transmission process for vaccinated people. Combined with the global mobility network reconstructed from the OAG international air traffic records, we successfully reproduce real-world COVID-19 epidemic dynamics with $R^2 > 0.990$ on both global and regional scales. Our model provides a solid real-world basis for evaluating practical vaccine sharing strategies, which is an important advantage compared with existing research. As an example, Wagner et al. [14] construct a highly simplified scenario with only two regions without fitting to real-world epidemic, which forbids them to propose a practical vaccine sharing strategy that suits the real-world scenario with complicated international mobility. Similarly, Lampert et al. [13] propose a game theoretic approach to identify the condition for hypothetical vaccine-rich countries to share their extra vaccines, while the lack of epidemiological models weakens their credibility. Rotesi et al. [15] find that after reaching herd immunity, the US could donate vaccine doses to other countries without harming their own interest. However, the absence of real-world mobility patterns weakens its conclusion, and the absence of breakthrough infection processes makes it fails to provide concrete incentives for vaccine-rich countries. Going beyond these important theoretical works in hypothetical scenarios, our model enables the practical policy design rooted in strong real-world data.

Specifically, our model maps the vaccine distributions to their potential outcomes, supporting the quantitative evaluation of *enlightened self-interest* incentive mechanism. Based on the model, we showcase the feasibility of *enlightened self-interest* in COVID-19 vaccine sharing, where vaccine-producing regions can share their vaccines with other parts of the world after nearly 80% of their domestic people are vaccinated. Contradicting the common idea that sharing vaccines must sacrifice their own interest, we highlight that the vaccine sharing guided by the *enlightened self-interest* incentive mechanism can relieve epidemic outcomes in the vaccine-producing regions, which even surpass those in non-vaccine-producing regions. This phenomenon provides strong incentives for active vaccine sharing where the epidemic outcomes in both vaccine-producing regions and non-vaccine-producing regions can be simultaneously improved. We provide a larger feasible region for safely sharing vaccines with high credibility than existing works [13-15], where the global mobility enables quick action in promoting vaccine equity and the consideration of breakthrough infection guarantees conservative analyses. Furthermore, we demonstrate the mechanism behind such win-win vaccine sharing strategies, where the benefit from reduced internationally imported cases outweighs the threat from increased local infections due to the disproportionate vaccine effect and highly intertwined global mobility. Finally, we evaluate the benefit of active vaccine sharing in terms of global connectivity. While the strong travel restriction policies might be necessary to curb the pandemic, they also considerably disrupted global supply chains and undermined the world economy [37,38]. However, simply lifting these policies might lead to unexpected epidemic resurgence. By more actively sharing vaccine resources under *enlightened self-interest* incentive mechanism, up to 94.8% (95% CI: 94.3%-95.3%, $P<0.001$) of inter-regional mobility reduction can be averted compared with the *selfish* strategy, promoting the free movements of talents and goods [39-41] in these regions. Under a more equitable global vaccine distribution, vaccine-producing regions can maintain a higher level of inter-regional mobility, providing another incentive for actively sharing vaccines.

Leveraging the proposed modelling approach, several different allocation problems can be solved to promote more equitable resource distribution driven by *enlightened self-interest*. Specifically, we summarize three different scenarios that can be solved by the proposed model with necessary adaptions. First, the medicines.

Similar to vaccine resources for COVID-19, equitably distributing drugs for other infectious diseases is also critical to mitigating the possible global crisis, such as ZMapp for Ebola [42]. Considering the similarity in cost and efficacy of vaccines and medicines, our model can be extended to the medicine allocation problem with few modifications. Second, the medical equipment. Different from vaccines and medicines that usually have a high research and development cost but a low production cost, medical equipment has a high price and is not intended for one-time use. Besides, the benefit of medical equipment is not to reduce the possible transmission, but to resecure severe patients. We can adapt the epidemiological model to capture the imbalance between patient demand and the availability of medical equipment in terms of time and space, which further informs the years of potential life lost (YPLL) or human capital metrics [43] to solve the allocation problem from the *enlightened self-interest* perspective. Third, the health workforce. By organizing international medical teams and flexibly deploying the health workforce, the potential impact of epidemic outbreaks can be greatly mitigated [44]. Unlike the distribution of high-value medical equipment, the cost of scheduling healthcare personnel is much lower. However, due to the long training procedure required for skilled healthcare professionals, the cost of potential loss is extremely high. With proper modelling approaches to quantify the cost and benefit, we can deploy the proposed model in a series of allocation tasks.

## Limitations

There are several limitations in the present work. First, our proposed model adopts a classic compartmental framework, which might simplify real-world epidemic dynamics. The primary source of uncertainties lies in the estimation of epidemiological parameters, *i.e.* the infection rate and recovery rate of COVID-19. To capture these uncertainties, our solution is three-fold: 1) We adopt a metapopulation framework that fits different sets of parameters for each region, accounting for the spatial heterogeneity. 2) We divide the simulation period into three phases according to the mutation timeline (see Supplementary Table **S6** in the Online Supplementary Document), accounting for the temporal heterogeneity. 3) We capture the randomness of the parameters by 40 random initializations of the model, which generates confidential intervals in all the analyses. However, it still relies on several classic assumptions. As an example, the countries within each region in the same phase share the same epidemiological parameters, and the vaccine effectiveness is constant across time. Besides, we also assume the epidemiological parameters are decoupled from different vaccine sharing strategies. More fine-grained behaviour and demographic data are needed to challenge these conventional assumptions. Since our model is proven effective in tracing global epidemic development, we leave the more fine-grained epidemic model as a future work.

Second, we adopt a relatively straightforward strategy in redistributing the shared vaccine doses according to regional population, without considering the capacity of medical infrastructures. However, this redistributing strategy is in accordance with the COVAX project, which is easier to be implemented in the real world. Nevertheless, with such a redistribution scheme we are still able to reveal the feasibility of a more equitable vaccine distribution, which in turn emphasizes the robustness of our findings. Our study provides general evidence that we can achieve such Pareto improvement and a more equitable vaccine distribution, based on which more detailed, fine-grained vaccine distribution can be explored by introducing more advanced modelling assumptions. To further magnify the benefit, future studies are needed to devise more targeted redistribution schemes based on the epidemic dynamics and detailed profiles of each region.

Third, we focus on how to share vaccines globally, without probing into the domestic distribution of vaccines. For domestic distribution, several studies have provided important insight [17,19,45]. However, the global vaccine distribution received woefully inadequate research attention due to the difficulty of reconstructing the worldwide pandemic, even though this problem is becoming increasingly important. With the enlightenment of *enlightened self-interest*, we have proved both the possibility and the importance of equitable vaccine sharing at the global level, and it is worth putting future efforts to combine these two perspectives of vaccine sharing. With simultaneous consideration of both domestic and global vaccine distribution, we can form a hierarchical vaccine distribution framework, providing more fine-grained insights for vaccine sharing.

Fourth, we noticed the possible reliability issue of COVID-19 case statistics during the early outbreaks of COVID-19 [46,47]. To minimize these errors, we refer to the most comprehensive systematic review of COVID-19 seroprevalence, which contains 968 studies with more than 9.3 million participants in 74 countries [48]. It helps to preclude a large percentage of studies with insufficient reliability (496, 51% of all the studies) and to provide cross-validated, corrected median seroprevalence data. Nevertheless, there may still be certain flaws in the case estimation during the early stage of the pandemic.

## CONCLUSIONS

In summary, we propose an accurate epidemiological model that simultaneously considers global vaccine distribution and human mobility, acting as a powerful tool for us to examine the feasibility and effectiveness of global vaccine sharing. Based on the proposed model, we highlight the surprising fact that the *enlightened self-interests* of high-income countries are aligned with more equitable global vaccine distribution. During the COVID-19 pandemic, we highlight that high-income countries should start sharing their extra vaccines after when nearly 80% of their citizens are vaccinated. To promote vaccine equity, the best weapon is not to criticize from the moral high ground, but to provide tangible incentives for the noble action. Our findings enlighten such incentives to promote multilateral collaborations in global vaccine redistribution and vaccine equity, which facilitates a mutually benefiting world. After all, in the combat against the pandemic, "no one is safe until everyone is safe" [49].

**Acknowledgements:** We thank Katherine Budeski for the careful proof-reading thought out the paper, which provides valuable language support from a native-speaker perspective.
**Data availability:** Most of the data used in our work can be freely accessed. The empirical epidemic data and vaccination data are available in the GitHub repository maintained by Johns Hopkins University [50] in https://github.com/CSSEGISandData/COVID-19. The global population data is maintained by World Bank in https://data.worldbank.org/indicator/SP.POP.TOTL. We purchase the global air traffic records from the Official Aviation Guide (OAG). It contains all the international flight orders at country level from January 2019 to July 2021. The data purchase agreement with OAG prohibits us from sharing these data with third parties, but interested parties can contact OAG to make the same data purchase. For the detailed description about the data in this study, please refer to the **Online Supplementary Document.**

**Funding:** This work was supported in part by The National Key Research and Development Program of China under grant 2020AAA0106000, the National Natural Science Foundation of China under U1936217.

**Authorship contributions:** Y.L., K.T. and F.X. contributed to conceptualizing the study. Z.H., C.L., QY.H. processed the raw data, performed the experiments, and prepared all the figures. Q.H., K.B., D.J. provided critical revisions. All the authors jointly analysed the results and participated in writing the manuscript.

**Disclosure of interest:** The authors completed the ICMJE Disclosure of Interest Form (available upon request from the corresponding author) and disclose no relevant interests.

Additional material
**Online Supplementary Document**.

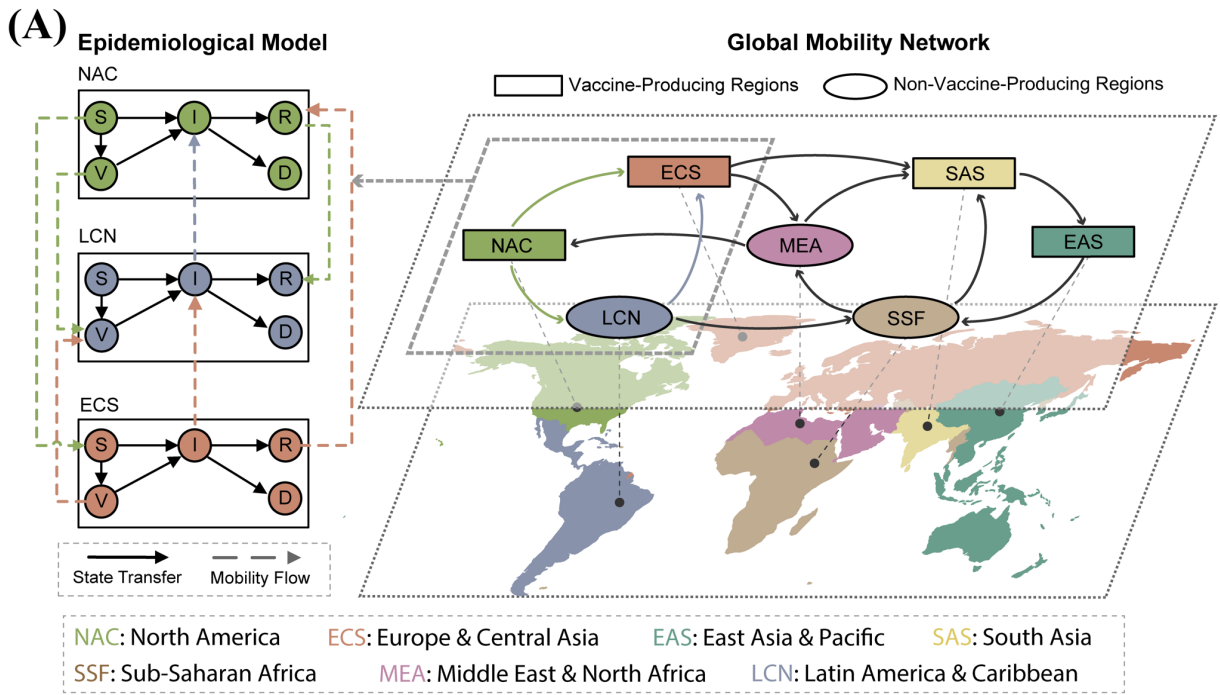

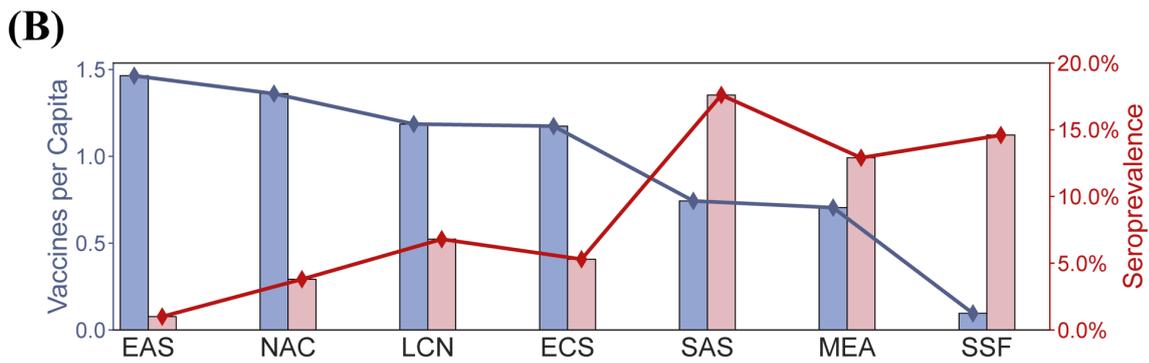

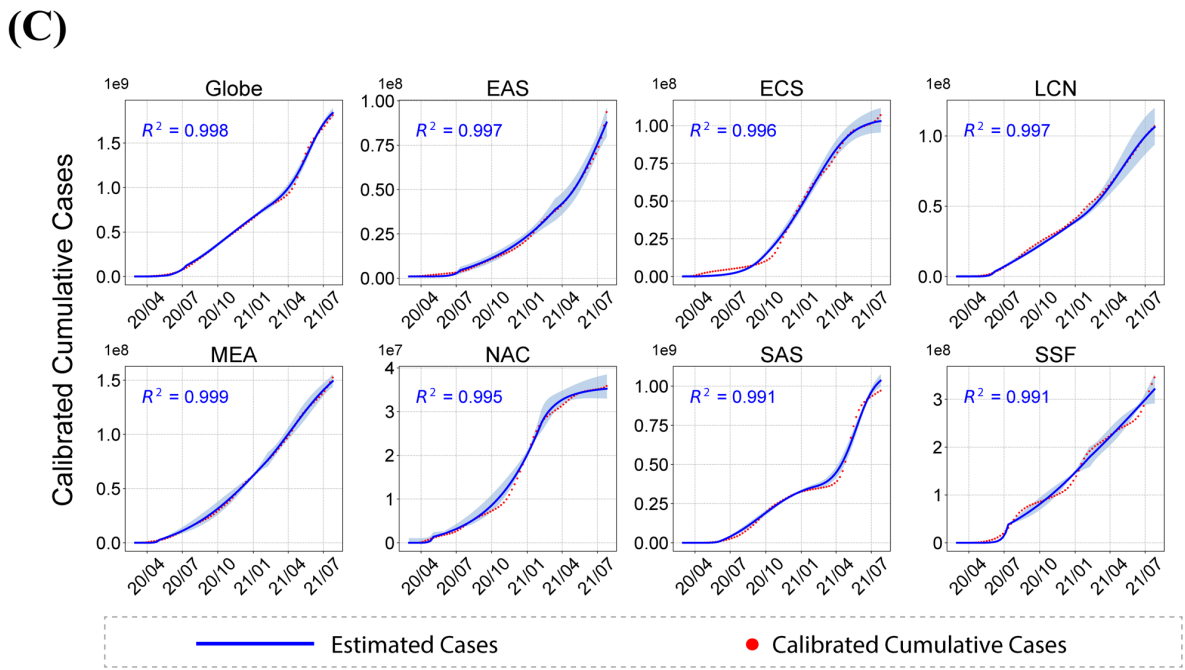

**Figure 1.** Model description and its performance in capturing real-world epidemic dynamics.

(A) Illustration of our model framework. The globe is divided into 7 geographical regions, where each region consists of an individual epidemiological model considering breakthrough infections after vaccinated. People in different regions can move from one region to another, forming a global mobility network that characterized by the air traffic data from OAG. (B) Vaccines per capita and seroprevalence of COVID-19 in each region until 2021-11-14. Vaccine availability disobeys the epidemic situation in different regions. Specifically, SAS has the highest seroprevalence, while its vaccines per capita are much less than high-income regions such as NAC. (C) Result of model fitting in global and regional level, where shaded area represents the 99% confidence interval. We have $R^2 > 0.990$ for all the regions, demonstrating the accuracy of our proposed model.

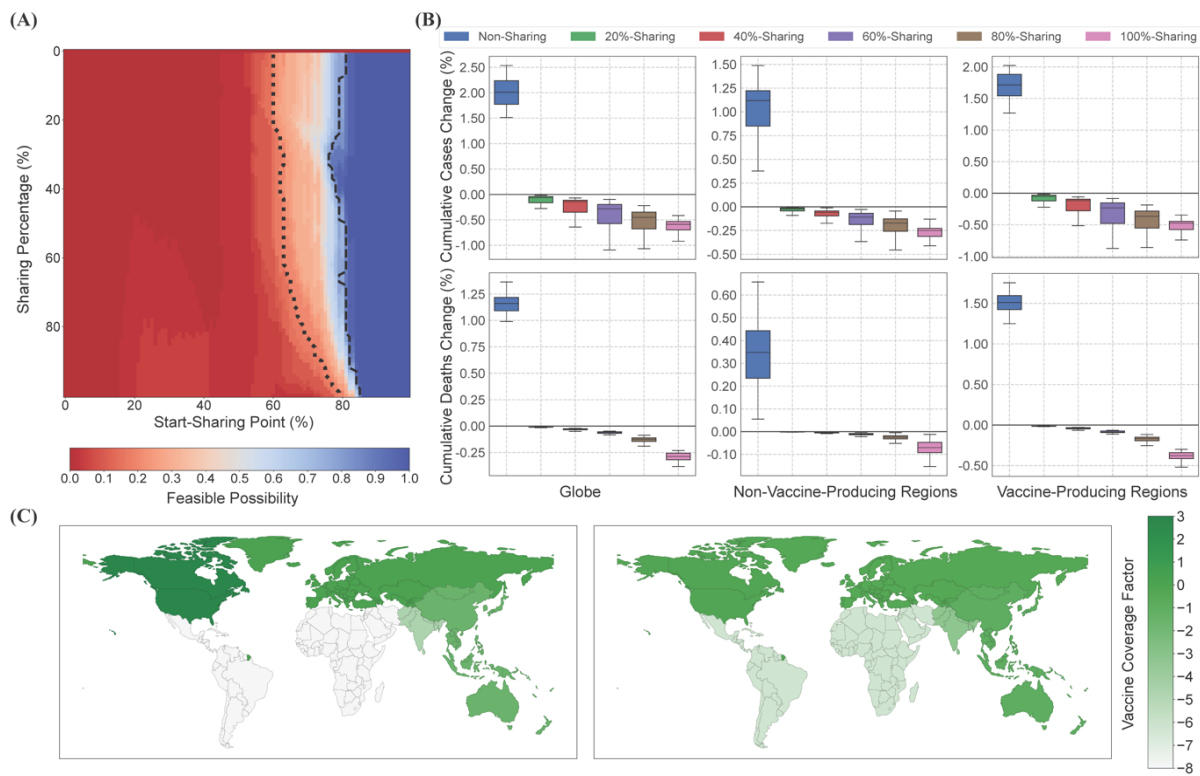

**Figure 2.** Impacts of enlightened self-interest guided vaccine sharing strategies.

(A) Feasible parameter combinations of enlightened self-interest guided vaccine sharing strategies. The colours indicate the possibilities for each vaccine-producing region and the globe to have fewer infections compared with selfish strategy. Strategies that can achieve such simultaneous improvement over 80% possibility are noted as enlightened self-interest incentive mechanism. The boundary of 20% and 80% possibility are labelled by the dotted line and dashed line accordingly. (B) Increase of infections and deaths for different vaccine sharing strategies compared with selfish strategy around the globe, in the non-vaccine-producing regions and in the vaccine-producing regions. (C) Geographical vaccine distribution under the selfish strategy (left) and the 100%-sharing strategy (right). The vaccine coverage factor is calculated by the logarithm of the ratio between the number of vaccines and the population.

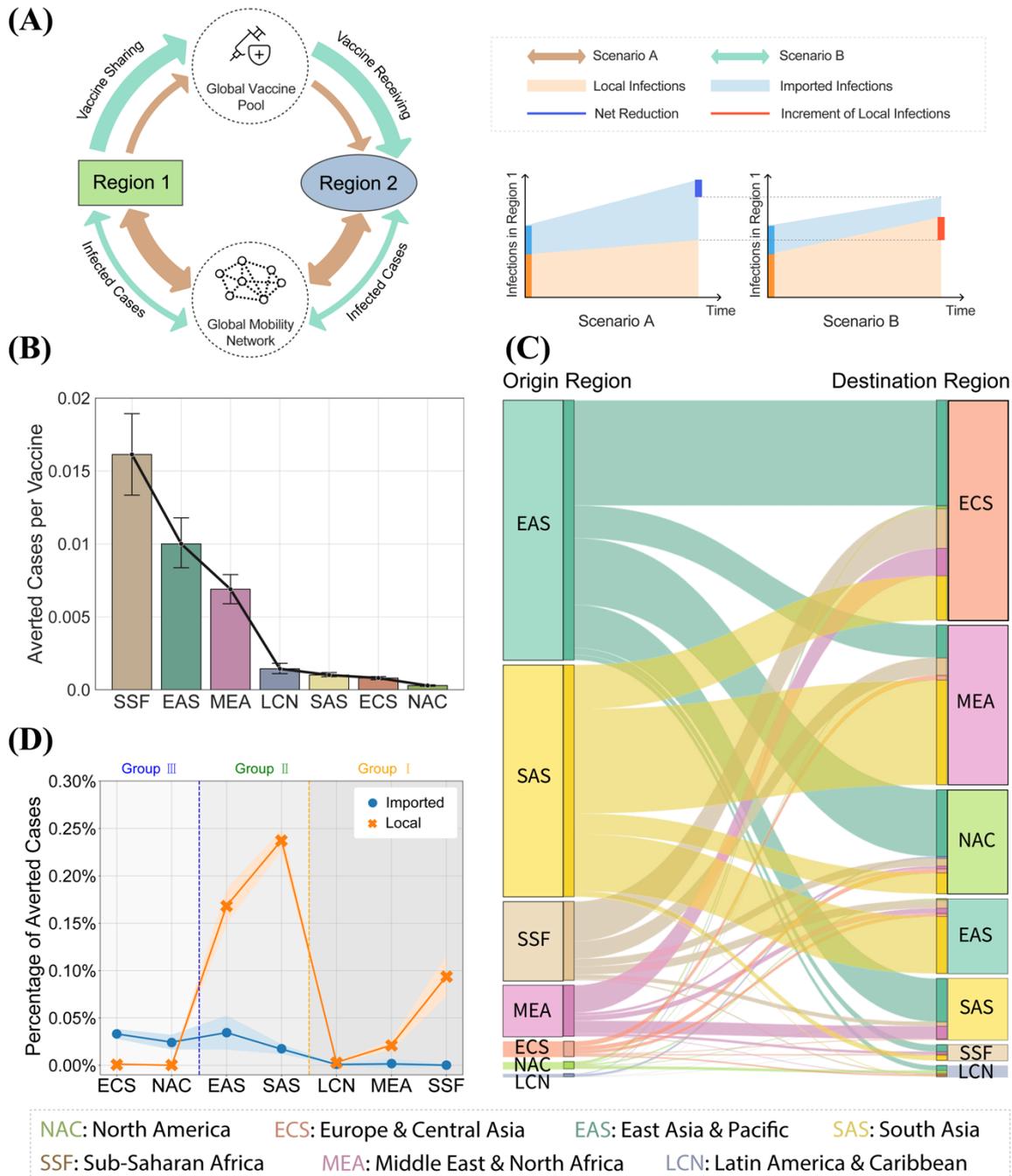

**Figure 3.** Rationalizing the enlightened self-interest incentive mechanism.

(A) Illustration of the incentive mechanism behind the enlightened self-interest guided vaccine sharing strategies. By categorizing the infection into local infections (orange area) and imported infections (blue area), we expect under more active vaccine sharing strategies, there will be a net reduction in infections (deep blue line), although the local infections may increase (red line), due to the reduced imported infections. (B) Marginal utility of vaccines based on real-world vaccine distribution. It demonstrates that the effectiveness of extra vaccines vanishes in regions with high vaccination rates, such as NAC and ECS. The error bar demonstrates 95% of confidence intervals. (C) Flow of reduced infection cases under the enlightened self-interest guided vaccine sharing strategy, started from 90% vaccination with 90% vaccines shared, comparing with the selfish strategy. (D) Quantifying the disentangled imported transmission and local transmission. For Group Ⅰ, they have reduced local transmission due to vaccine sharing. For Group Ⅱ, both the imported transmission and local transmission will reduce. For Group Ⅲ, the received imported transmission benefit outweighs the increasing of local transmission. The shaded area represents the standard deviation.

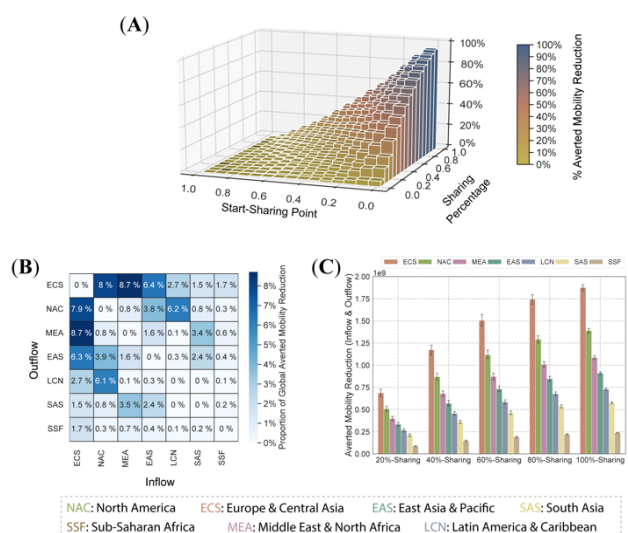

**Figure 4.** Active vaccine sharing promotes inter-region mobility.

(A) Percentage of averted reduction in mobility under different vaccine sharing strategies. A more active sharing can prevent more mobility reduction, where up to 94.8% (95% CI: 94.3%-95.3%, $P<0.001$) of mobility reduction can be averted. (B) Proportion of averted reduction in global mobility for each region, under the enlightened self-interest guided vaccine sharing strategy. (C) Volume of averted reduction in mobility under different enlightened self-interest guided vaccine sharing strategies, where error bars represent the 99% confidence interval. From (B) and (C), we can find vaccine-producing regions, such as ECS and NAC, are expected to benefit most from mobility recovery in both relative proportion and absolute mobility volume.

**Correspondence to:**

Fengli Xu

Department of Electronic Engineering, Tsinghua University, Beijing, P. R. China,
fenglixu@tsinghua.edu.cn

Kun Tang

Vanke School of Public Health, Tsinghua University, Beijing, P. R. China

Institute for Healthy China, Tsinghua University, Beijing, P. R. China

tangk@mail.tsinghua.edu.cn

Yong Li

Beijing National Research Center for Information Science and Technology (BNRist), Beijing, P. R. China

Department of Electronic Engineering, Tsinghua University, Beijing, P. R. China

liyong07@tsinghua.edu.cn